\documentclass[aps,prl,twocolumn,amsmath,amssymb,superscriptaddress]{revtex4}
\usepackage{graphicx}
\usepackage{amsmath}

\begin{document}

\title{Spin-Orbit Coupled Fermi Gases across a Feshbach Resonance}
\author{Zeng-Qiang Yu}
\email{zqyu.physics@gmail.com}
\affiliation{Institute for Advanced Study, Tsinghua University, Beijing, 100084, China}
\author{Hui Zhai}
\email{hzhai@mail.tsinghua.edu.cn}
\affiliation{Institute for Advanced Study, Tsinghua University, Beijing, 100084, China}
\date{\today}
\begin{abstract}
In this letter we study both ground state properties and the superfluid transition temperature of a spin-$1/2$ Fermi gas  across a Feshbach resonance with a synthetic spin-orbit coupling, using mean-field theory and exact solution of two-body problem. We show that a strong spin-orbit coupling can significantly enhance the pairing gap for $1/(k_{\text{F}}a_{\text{s}})\lesssim 0$ due to increased density-of-state. Strong spin-orbit coupling also significantly enhances the superfluid transition temperature when $1/(k_{\text{F}}a_{\text{s}})\lesssim 0$, while suppresses it slightly when $1/(k_{\text{F}}a_{\text{s}})\gg 0$. The universal interaction energy and pair size at resonance are also discussed.
\end{abstract}
\maketitle

During the last few years, studies of ultracold Fermi gases across a Feshbach resonance (FR) have brought a lot of excitements to physics \cite{Sandro}. On a separate development, recent experimental breakthrough on synthetic gauge field has open up a lot of new opportunities to cold atom physics \cite{NIST1,NIST2}. One application of this technique is to engineer an effective spin-orbit coupling (SOC) in cold atom system \cite{Dalibard}. Very recently, a pioneer experiment in NIST has already achieved a restricted class of spin-orbit coupled BEC of $^{87}$Rb atoms \cite{NIST2}. Theoretically, even the mean-field study of boson condensate with SOC has revealed many interesting physics \cite{SObec,SObecHo,Yipetal}. For fermions, a concrete scheme has also been proposed for generating SOC in $^{40}$K atom in the regime where a magnetic FR is available \cite{fermion_proposal}, and the experiment of implementing this proposal is now going on in the laboratory. However, a theoretical study of spin-orbit coupled Fermi gases across a FR is still lacking.

In the absence of SOC, a Fermi gas across a FR possesses three key physical properties: i) across a FR the system undergoes a crossover from a BCS type fermion superfluid to a BEC of molecules; ii) at the FR, it is a strongly interacting system and exhibits many universal behaviors; iii) nearby a FR, the transition temperature of fermion superfluid $T_\text{c}/T_\text{F}$ is the highest one among all fermion superfluids (or superconductors). The question is that how these three properties evolve in the presence of SOC. For i), since now the pair wave-function has more complicated structure with both singlet and triplet components, and exhibits $p$-wave character in the helicity bases, one needs to investigate whether it is still a crossover or there is a phase transition in between. Even if it is still a crossover, how SOC affects it. For ii), since the strength of SOC introduces another length scale $\lambda/k_{\text{F}}$, the universal constants at resonance now become universal functions of $\lambda/k_{\text{F}}$, and we want to understand the behaviors of these functions. And for iii), the question is whether $T_{\text{c}}/T_{\text{F}}$ will be enhanced or suppressed by SOC. (Here the units $k_{\text{F}}$, $E_{\text{F}}$ and $T_{\text{F}}$ are the Fermi momentum, the Fermi energy and the Fermi temperature for non-interacting system without SOC. $a_{\text{s}}$ is the $s$-wave scattering length.)

In this letter we address these issues using both mean-field (MF) theory and exact solution of two-body (TB) problem, and the main results are summarized as follows:

(\textbf{A1}) The system is gapped for all values of $a_{\text{s}}$. The pair wave-function obtained from MF theory has the same symmetry property as the wave-function of TB bound state, and they coincide with each other for $1/(k_{\text{F}}a_{\text{s}})\gg 1$. These two evidences support a crossover picture instead of a phase transition.

(\textbf{A2}) The order parameter $\Delta$ always increases as the strength of SOC $\lambda/k_{\text{F}}$ increases. For $1/(k_{\text{F}}a_{\text{s}})\lesssim 0$, the increasing becomes profound when $\lambda/k_{\text{F}}$ is large enough that the density-of-state (DOS) at Fermi surface is significantly enhanced. While for $1/(k_{\text{F}}a_{\text{s}})\gg 0$ the increasing of $\Delta$ is always less significant when the chemical potential drops below the single particle energy minimum.

(\textbf{B}) At resonance, the interaction energy $E_{\text{int}}/E_{\text{F}}$ and the pair size $k_{\text{F}}l$ as functions of $\lambda/k_{\text{F}}$ have very different behaviors for $\lambda/k_{\text{F}}\ll1$ or $\gg 1$.

(\textbf{C1}) For $1/(k_{\text{F}}a_{\text{s}})< 0$, $T_{\text{c}}$ is enhanced by SOC due to two effects. One is that the increased DOS enhances $T_{\text{BCS}}$ from MF theory; and the other is because stable molecules with finite binding energy now also exist in this regime.

(\textbf{C2}) For $1/(k_{\text{F}}a_{\text{s}})\gg 0$, $T_{\text{c}}$ is given by BEC temperature of molecules $T_{\text{BEC}}$, and is slightly suppressed by SOC because the effective mass of molecules is increased.

(\textbf{C3}) At resonance $a_{\text{s}}=\pm\infty$, $T_{\text{c}}$ will finally saturate to $0.193T_{\text{F}}$ when $\lambda/k_{\text{F}}$ is large enough, which is higher than the transition temperature without SOC.


{\it Model:} We consider an isotropic in-plane SOC. The single particle Hamiltonian is given by $\hat{H}_0={\bf p}^2/(2m)+\lambda{\bf p}_{\perp}\cdot\boldsymbol{\sigma}_{\perp}/m$, where ${\bf p}_{\perp}=(p_x,p_y)$ and $\sigma_{\perp}=(\sigma_x,\sigma_y)$ (set $\lambda>0$ and $\hbar=1$). The generalization to anisotropic and more complicated SOC is quite straightforward. In the second quantized form, $\mathcal{\hat{H}}_0 =\sum_{\bf p}[\epsilon_{{\bf p}}(c^\dag_{{\bf p}\uparrow}c_{{\bf p}\uparrow}+c^\dag_{{\bf p}\downarrow}c_{{\bf p}\downarrow})+\lambda p_{\perp} (e^{-i\varphi_{{\bf p}}}c^\dag_{{\bf p}\uparrow}c_{{\bf p}\downarrow}+e^{i\varphi_{{\bf p}}}c^\dag_{{\bf p}\downarrow}c_{{\bf p}\uparrow})]$, where $\epsilon_{{\bf p}}=p^2/(2m)$, $p_{\perp}=|{\bf p}_{\perp}|$ and $\varphi_{{\bf p}}=\text{arg}(p_x+ip_y)$. The single particle Hamiltonian can be diagonalized in the helicity bases as $
\mathcal{\hat{H}}_0=\sum_{{\bf p}}[\xi_{{\bf p}+}h^\dag_{{\bf p},+}h_{{\bf p},+}+\xi_{{\bf p}-}h^\dag_{{\bf p},-}h_{{\bf p},-}]$,
with $\xi_{{\bf p}\pm}=\epsilon_{{\bf p}}\pm\lambda p_{\perp}/m$, where helicity $\pm$ means that the in-plane spin is parallel or anti-parallel to the in-plane momentum. The fermion operators in the helicity bases are related to the fermion operators in the original spin bases via $h_{{\bf p},+}=(c_{{\bf p}\uparrow}+e^{-i\varphi_{{\bf p}}}c_{{\bf p}\downarrow})/\sqrt{2}$ and $h_{{\bf p},-}=(e^{i\varphi_{{\bf p}}}c_{{\bf p}\uparrow}-c_{{\bf p}\downarrow})/\sqrt{2}$.

When the effective range $r_0$ of inter-atomic potential is much smaller than all the other length scales in the problem, i.e. $k_{\text{F}}r_0\ll 1$ and $\lambda r_0\ll 1$, as in the conventional crossover theory, we use a zero-range potential to describe the intereaction between atoms. The interaction can be written as $\mathcal{\hat{H}}_{\text{int}}=(g/V)\sum_{{\bf p}{\bf p^\prime}{\bf q}}c^\dag_{{\bf q/2+p}\uparrow}c^\dag_{{\bf q/2-p}\downarrow}c_{{\bf q/2-p^\prime}\downarrow}c_{{\bf q/2+p^\prime}\uparrow}$, where $g$ is related to $a_{\text{s}}$ via $1/g=m/(4\pi a_{\text{s}})-\sum_{{\bf k}}1/(2\epsilon_{{\bf k}})$, and $V$ is the system volume.

\begin{figure}[tbp]
\includegraphics[width=3.1 in]
{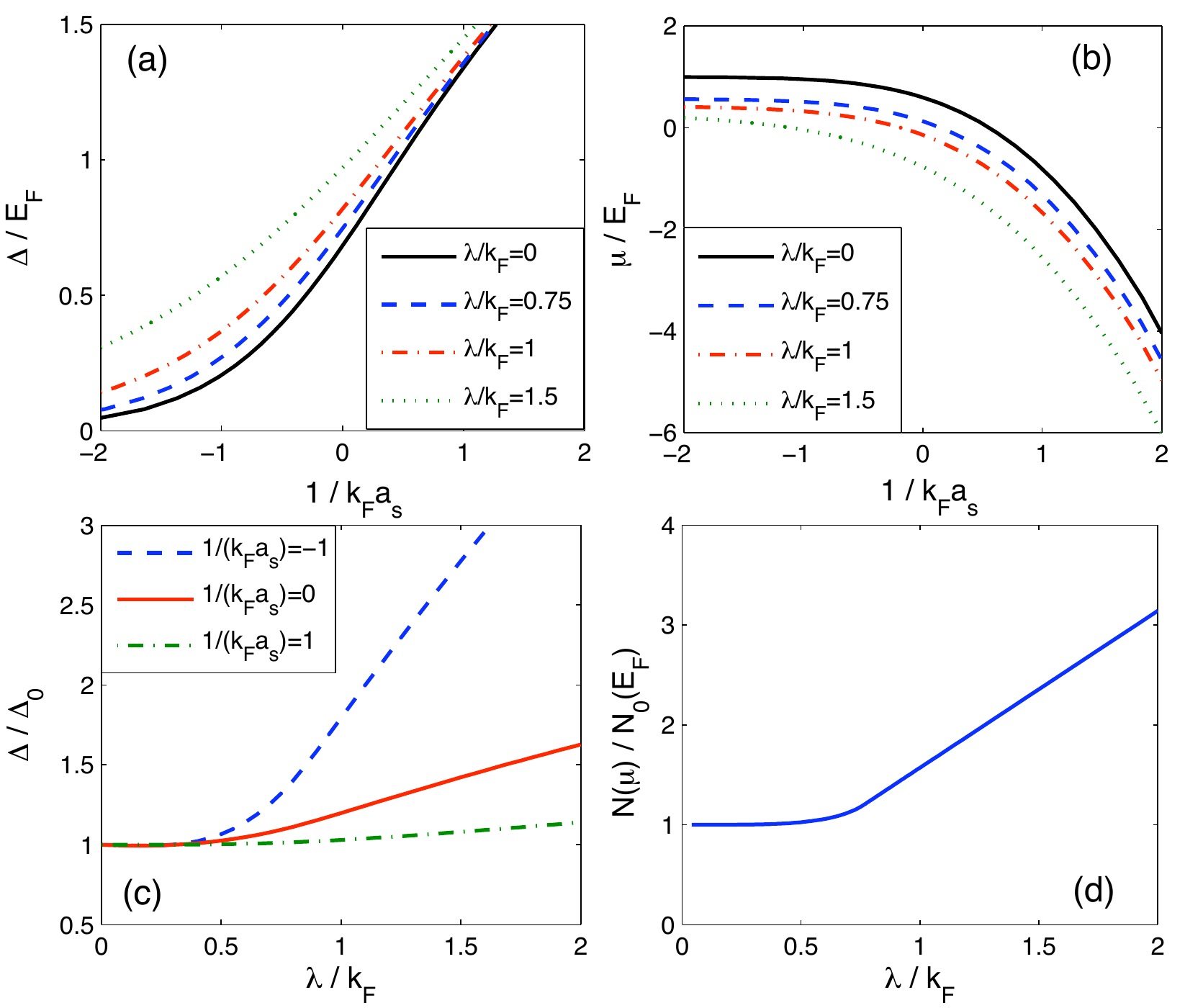}
\caption{(a,b) Order parameter $\Delta/E_{\text{F}}$ and chemical potential $\mu/E_{\text{F}}$ (measured from single particle minimum) as a function of $1/(k_{\text{F}}a_{\text{s}})$ for different $\lambda/k_{\text{F}}$; (c) $\Delta/\Delta_0$ as a function of $\lambda/k_{\text{F}}$ for three different values of $1/(k_{\text{F}}a_{\text{s}})$, where $\Delta_0$ is the paring gap without SOC. (d) DOS at Fermi energy $\mathcal{N}(\mu)$ compared to DOS without SOC ($\mathcal{N}_0(E_{\text{F}})$) as a function of $\lambda/k_{\text{F}}$. \label{Delta}}
\end{figure}

{\it Mean-field Theory:} For MF discussion, we only focus on ${\bf q}=0$ channel of $\mathcal{\hat{H}}_{\text{int}}$. Introducing the order parameter $\Delta=(g/V)\sum_{{\bf p}}\langle c_{{\bf -p}\downarrow}c_{{\bf p}\uparrow}\rangle$, one can obtain the mean-field interaction $\mathcal{\hat{H}}_{\text{int}}^{\text{MF}}=\Delta\sum_{{\bf p}}(c^\dag_{{\bf p}\uparrow}c^\dag_{{\bf -p}\downarrow}+\text{h.c.})-|\Delta|^2V/g$. Transforming it into the helicity bases, it becomes $
\mathcal{\hat{H}}_{\text{int}}^{\text{MF}}=-(\Delta/2)\sum_{{\bf p}}(e^{-i\varphi_{{\bf p}}}h^\dag_{{\bf p},+}h^\dag_{{\bf -p},+}+e^{i\varphi_{{\bf p}}}h^\dag_{{\bf p},-}h^\dag_{{\bf -p},-})-|\Delta|^2V/g$.
One can see that pairing only exits between atoms with same helicity, and the pairing of helicity $\pm$ has $p_x\mp ip_y$ symmetry.

Hence, the MF Hamiltonian is given by $\mathcal{\hat{H}}_{\text{MF}}=\mathcal{\hat{H}}_0+\mathcal{\hat{H}}^{\text {MF}}_{\text{int}}-\mu\hat{N}$. It is very to solve $\mathcal{\hat{H}}_{\text{MF}}$ in the helicity bases, which gives
\begin{align}
-\frac{m}{4\pi a_{\text{s}}}&=\frac{1}{4V}\sum\limits_{{\bf p}}\left[\frac{f_{{\bf p},+}}{\varepsilon_{{\bf p},+}}+\frac{f_{{\bf p},-}}{\varepsilon_{{\bf p},-}}-\frac{2}{\epsilon_{{\bf p}}}\right],\label{gapE}\\
n&=\frac{1}{V}\sum\limits_{{\bf p}}\left[1-\frac{\xi_{{\bf p},+}f_{{\bf p},+}}{2\varepsilon_{{\bf p},+}}-\frac{\xi_{{\bf p},-}f_{{\bf p},-}}{2\varepsilon_{{\bf p},-}}\right],\label{numE}
\end{align}
where $\varepsilon_{{\bf p}\pm}=\sqrt{(\xi_{{\bf p}\pm}-\mu)^2+\Delta^2}$ is the energy of quasi-particles and $f_{{\bf p},\pm}=\tanh[\varepsilon_{{\bf p}\pm}/(2k_{\text{B}}T)]$.

{\it Two-body Problem:} The TB problem in the presence of SO coupling has been solved in Ref. \cite{Vijay} for the case of molecular center-of-mass momentum ${\bf q}=0$. It was found that the TB bound state appears even at the BCS side of resonance with $a_{\text{s}}<0$, because of the increase of low-energy DOS \cite{Vijay}. Here we solve the two-body problem for finite ${\bf  q}$, which is very useful for later discussions. In general, the TB wave-function can be assumed as $|\Psi\rangle_{{\bf q}}=\sum_{{\bf k}}'[\psi_{\uparrow\downarrow}({\bf k})c^\dag_{{\bf q/2+k}\uparrow}c^\dag_{{\bf q/2-k}\downarrow}+\psi_{\downarrow\uparrow}({\bf k})c^\dag_{{\bf q/2+k}\downarrow}c^\dag_{{\bf q/2-k}\uparrow}+\psi_{\uparrow\uparrow}({\bf k})c^\dag_{{\bf q/2+k}\uparrow}c^\dag_{{\bf q/2-k}\uparrow}+\psi_{\downarrow\downarrow}({\bf k})c^\dag_{{\bf q/2+k}\downarrow}c^\dag_{{\bf q/2-k}\downarrow}]$, where $\sum'$ means the summation is over half of momentum space. The Schr\"odinger equation $(\mathcal{\hat{H}}_0+\mathcal{\hat{H}}_{\text{int}})|\Psi\rangle_{{\bf q}}=E_{{\bf q}}|\Psi\rangle_{{\bf q}}$ leads to a self-consistency equation as \cite{Supplementray}
\begin{equation}
\frac{m}{4\pi a_{\text{s}}}=\sum\limits_{{\bf k}}\frac{\mathcal{E}_{{\bf k},{\bf q}}}{\mathcal{E}_{{\bf k},{\bf q}}^2-\frac{4\lambda^2 k^2_{\perp}}{m^2}-\frac{4\lambda^4 k^2_{\perp}q^2_{\perp}\sin^2\varphi_{{\bf k}{\bf q}}}{m^2(m^2\mathcal{E}_{{\bf k},{\bf q}}^2-\lambda^2 q_{\perp}^2)}}+\frac{1}{2\epsilon_{{\bf k}}}, \label{two-body}
\end{equation}
where $\mathcal{E}_{{\bf k},{\bf q}}=E_{{\bf q}}-\epsilon_{{\bf q/2+k}}-\epsilon_{{\bf q/2-k}}$, and $\varphi_{{\bf k}{\bf q}}=\varphi_{{\bf k}}-\varphi_{{\bf q}}$. For ${\bf q}=0$, Eq. (\ref{two-body}) recovers the results in Ref. \cite{Vijay}, and for any $a_{\text{s}}$ there is always a bound state solution $E_{0}<-\lambda^2/m$. We obtain an analytical equation for $E_0$
\begin{equation}
\frac{2}{a_{\text{s}}}=2\sqrt{(-E_0)m}-\lambda\ln\frac{\sqrt{(-E_0)m}+\lambda}{\sqrt{(-E_0)m}-\lambda}.\label{E_0}
\end{equation}
With $E_{0}$, one can then use the Schr\"odinger equation to determine the bound state wave-function.

\begin{figure}[bp]
\includegraphics[height=1.5in, width=3.4 in]
{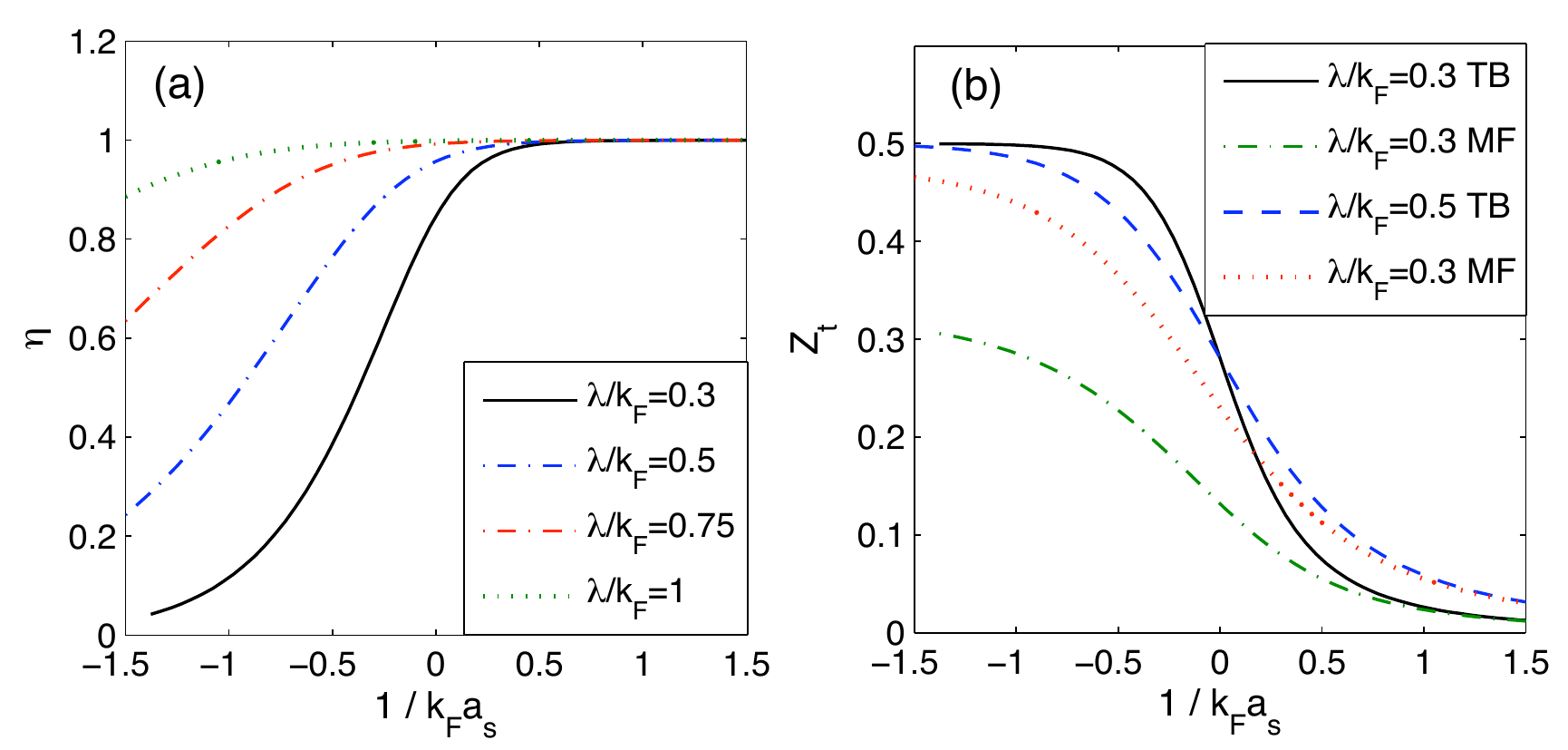}
\caption{(a) $\eta=\langle\Phi|\Psi\rangle_{{\bf q}=0}$ as a function of $1/(k_{\text{F}}a_{\text{s}})$ for different $\lambda/k_{\text{F}}$. (b) The ratio of triplet component to singlet component $Z_{\text{t}}$ for TB wave-function and MF pair wave-function, for two different $\lambda/k_{\text{F}}$. \label{wf}}
\end{figure}

{\it Results and Discussions:} With the MF theory and TB solution presented above, we are ready to address the questions posted at the beginning.

{\bf (A) Crossover:} We solve the MF equation for $T=0$. The value of order parameter $\Delta/E_{\text{F}}$ and chemical potential $\mu/E_{\text{F}}$ as a function of $1/(k_{\text{F}}a_{\text{s}})$ is shown in Fig. \ref{Delta}(a-b), for various $\lambda/k_{\text{F}}$. Not surprisingly, their behaviors are not qualitatively different from $\lambda=0$. Nevertheless, it is worth to point out that for conventional $p$-wave pairing, even though the order parameter is non-zero, the pairing gap will still close when the Fermi surface touches ${\bf p}_\perp=0$ point. However, the paring form factor here is in fact $(p_x\pm i p_y)/p_{\perp}$ instead of conventional $p_x\pm i p_y$, which ensures that the superfluid is always gapped.

From MF theory, we can obtain the BCS wave-function
\begin{equation}
|\text{BCS}\rangle\propto\exp\Big[{\sum_{{\bf k}}}'\frac{v_{{\bf k},+}}{u_{{\bf k},+}}h^\dag_{{\bf k},+}h^\dag_{{\bf -k},+}+\frac{v_{{\bf k},-}}{u_{{\bf k},-}}h^\dag_{{\bf k},-}h^\dag_{{\bf -k},-}\Big]|0\rangle,\nonumber
\end{equation}
where $v_{{\bf k},\pm}=e^{\mp i\varphi_{{\bf k}}}\sqrt{{1\over 2}\big(1-{\xi_{{\bf k},\pm}-\mu\over \varepsilon_{{\bf k},\pm}}\big)}$ and $u_{{\bf k},\pm}=\sqrt{{1\over 2}\big(1+{\xi_{{\bf k},\pm}-\mu\over \varepsilon_{{\bf k},\pm}}\big)}$.
Then we can define a pair wave-function as $|\Phi\rangle=\sum_{{\bf k}}'\big[\phi_{\uparrow\downarrow}({\bf k})c^\dag_{{\bf k}\uparrow}c^\dag_{{\bf -k}\downarrow}+\phi_{\downarrow\uparrow}({\bf k})c^\dag_{{\bf k}\downarrow}c^\dag_{{\bf -k}\uparrow}+\phi_{\uparrow\uparrow}({\bf k})c^\dag_{{\bf k}\uparrow}c^\dag_{{\bf -k}\uparrow}+\phi_{\downarrow\downarrow}({\bf k})c^\dag_{{\bf k}\downarrow}c^\dag_{{\bf -k}\downarrow}\big]|0\rangle$, where
\begin{align}
\phi_{\uparrow\downarrow}({\bf k})&=-\phi_{\downarrow\uparrow}({\bf k})={-1\over \sqrt{\mathcal{C}}}\left(\Big|\frac{v_{{\bf k},+}}{u_{{\bf k},+}}\Big|+\Big|\frac{v_{{\bf k},-}}{u_{{\bf k},-}}\Big|\right), \\
\phi_{\uparrow\uparrow}({\bf k})&=-\phi_{\downarrow\downarrow}^*({\bf k})=\frac{e^{-i\varphi_{{\bf k}}}}{\sqrt{\mathcal{C}}}\left(\Big|\frac{v_{{\bf k},+}}{u_{{\bf k},+}}\Big|-\Big|\frac{v_{{\bf k},-}}{u_{{\bf k},-}}\Big|\right),
\end{align}
and $\mathcal{C}$ is the normalization factor. The symmetry properties of pair wave-function $|\Phi\rangle$
agree with that of zero-momentum molecular wave-function $|\Psi\rangle_{{\bf q}=0}$ discussed in Ref. \cite{Vijay}.
We then compute their overlap $\eta=\langle \Phi|\Psi\rangle_{{\bf q}=0}$ as a function of $1/(k_{\text{F}}a_{\text{s}})$, as shown in Fig. \ref{wf}(a). The overlap approaches unity rapidly when $1/(k_{\text{F}}a_{\text{s}})\gtrsim 0$. In Fig. \ref{wf}(b) we also plot the ratio of triplet to singlet component for both $|\Phi\rangle$ and $|\Psi\rangle_{{\bf q}=0}$. It shows that the TB wave-function always has a larger triplet component. Nevertheless, they converge together quickly.

In Fig. \ref{Delta}(c) we plot $\Delta$ as a function of $\lambda/k_{\text{F}}$, from which one can see that there is a characteristic value roughly located at $\lambda/k_{\text{F}}\approx 0.5$. Below this value the change of $\Delta$ with $\lambda/k_{\text{F}}$ is small, while above this value the increasing of $\Delta$ becomes very significant.  In Fig \ref{Delta}(d) we show the DOS at Fermi energy $\mathcal{N}(\mu)$ compared to the DOS without SOC ($\mathcal{N}_0(E_{\text{F}})$). Their ratio remains nearly unity until reaching $\lambda/k_{\text{F}}\approx  0.5$, and then it increases rapidly. It is because for low density or strong SOC, the Fermi energy drops below Dirac point at ${\bf p}_{\perp}=0$ with $k_{\text{F}}<(3\pi/4)^{1/3}\lambda$, and only the lower helicity minus branch will be occupied. In this case its DOS $\mathcal{N}(\xi)=m\lambda/(2\pi)$ is a constant independent of $\xi$, while without SOC, $N(\xi)\sim\sqrt{\xi}$, therefore the DOS is always increased by SOC \cite{Supplementray}.

By comparing Fig. \ref{Delta}(c) and (d) one can draw the conclusion that the increasing of $\Delta$ is due to the increasing of DOS. From Fig. \ref{Delta}(a,b), we also notice that when $\mu$ decreases below the single particle energy minimum, the DOS effect is no longer important. Thus, the influence of SOC on $\Delta$ becomes very weak.

\begin{figure}[bp]
\includegraphics[height=1.5in, width=3.4 in]
{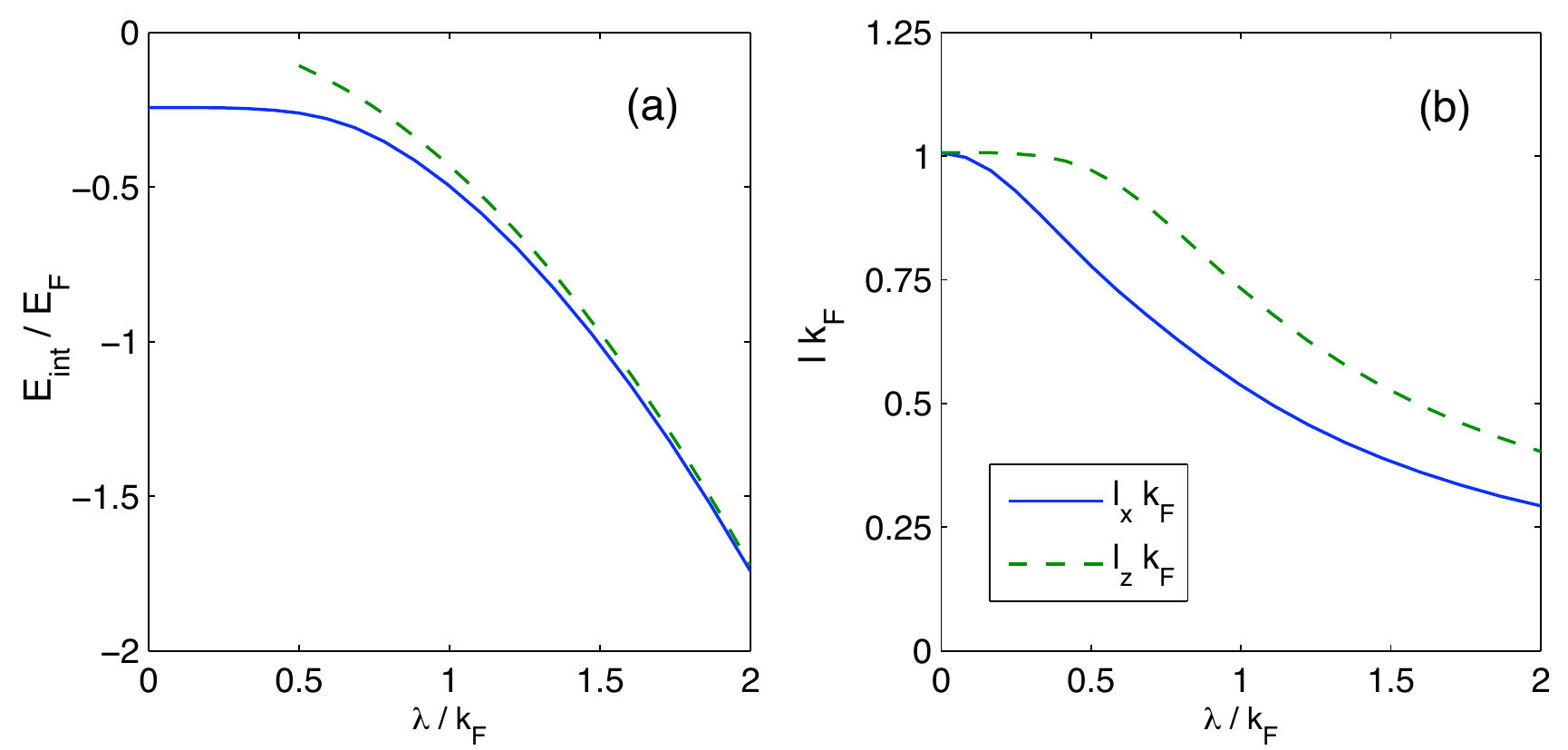}
\caption{(a) $E_{\text{int}}/E_{\text{F}}$ as a function of $\lambda/k_{\text{F}}$. The dashed line is a fit of $-0.44(\lambda/k_{\text{F}})^2$; (b) Size of Cooper pair in $x$-$y$ plane $l_{x}k_{\text{F}}$ and along $\hat{z}$ direction $l_{z}k_{\text{F}}$ as functions of $\lambda/k_{\text{F}}$. Both are plotted at resonance for $a_{\text{s}}=\infty$.  \label{resonance}}
\end{figure}

{\bf (B) Universality:} As we all know very well now, at resonance when $a_{\text{s}}\rightarrow \pm \infty$, the interaction energy per particle $E_{\text{int}}/E_{\text{F}}$ will not diverge, instead, it saturates to a universal value of the order of unity. Now, this universal value becomes a function of $\lambda/k_{\text{F}}$. Within MF theory, we define the interaction energy as $E_{\text{int}}=(\langle \text{BCS}|\mathcal{\hat{H}}_0+\mathcal{\hat{H}}_{\text{int}}|\text{BCS}\rangle-\mathbb{E}_0)/N$, where $\mathbb{E}_0$ is the total energy of a non-interacting system. In Fig. \ref{resonance}(a), we plot $E_{\text{int}}/E_{\text{F}}$ as a function of $\lambda/k_{\text{F}}$ at resonance. Its behavior is very different in the regime of small and large $\lambda/k_{\text{F}}$. For $\lambda\ll k_{\text{F}}$, we have $E_{\text{int}}/E_{\text{F}}\approx -0.24+o(\lambda/k_{\text{F}})$; while for $\lambda\gg k_{\text{F}}$, we find $E_{\text{int}}/E_{\text{F}}\approx-0.44 (\lambda/k_{\text{F}})^2$. It is because from Eq. (\ref{E_0}) one can find out that at resonance, TB bound state energy $E_0=-2.88\lambda^2/(2m)$, and the binding energy is given by $-\lambda^2/m-E_0=0.88 \lambda^2/(2m)$, which is twice of $-E_{\text{int}}$ in the limit of strong SOC.

Another notable feature of unitary regime is that the size of Copper pairs $k_{\text{F}}l$ is also of the order of unity. Here we can compute the anisotropic pair size from
\begin{equation}
l_{\alpha}=\sqrt{{\sum_{{\bf k}}}'\left[2\left|\nabla_{k_\alpha}\phi_{\uparrow\downarrow}\right|^2+|\nabla_{k_\alpha}\phi_{\uparrow\uparrow}|^2 +|\nabla_{k_\alpha}\phi_{\downarrow\downarrow}|^2\right]},\nonumber
\end{equation}
where $\alpha=x,y$ and $z$. When $\lambda/k_{\text{F}}\neq 0$, $l_x=l_y<l_z$ which means the Cooper pairs are elongated, as shown in Fig. \ref{resonance}(b). Similarly, we find for small $\lambda/k_{\text{F}}$, $k_{\text{F}}l\approx 1+o(\lambda/k_{\text{F}})$, while for large $\lambda/k_{\text{F}}$, $k_{\text{F}}l \propto k_{\text{F}}/\lambda$. Their behaviors at large $\lambda/k_{\text{F}}$ shows that the system still behaves like weakly interacting molecular BEC.

\begin{figure}[tbp]
\includegraphics[width=3in]
{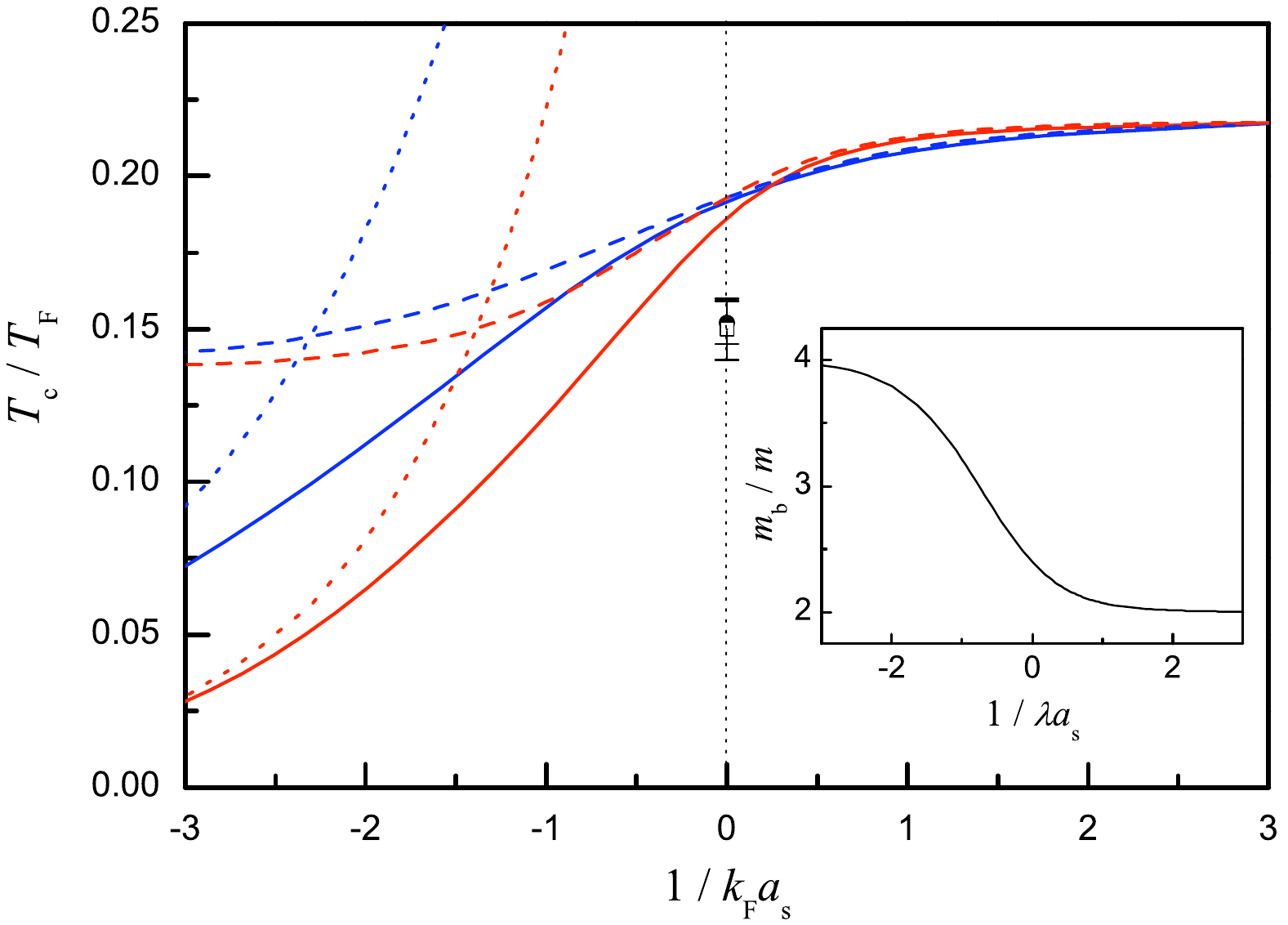}
\caption{Superfluid transition temperature $T_{\text{c}}/T_{\text{F}}$ from MF theory ($T_{\text{BCS}}$ dotted lines), from BEC temperature of molecules ($T_{\text{BEC}}$ dashed lines), and from an interpolation scheme including the contributions from non-condensed pairs (solid lines). The symbol with error bar is the Monte Carlo results without SOC \cite{Tc}. Inset: The in-plane effective mass of TB molecule $m_{\text{b}}/m$ as a function of $1/(\lambda a_{\text{s}})$.   \label{Tc}}
\end{figure}

{\bf (C) Superfuid Transition Temperature:} With MF theory, one can calculate the BCS temperature $T_{\text{BCS}}/T_{\text{F}}$ as shown in Fig. \ref{Tc}. It increases as $\lambda/k_{\text{F}}$ increases, for the same reason of DOS effect.

SOC also affects the effective mass of molecules. Substituting the molecular dispersion $E_{{\bf q}}=E_{0}+q^2_{\perp}/(2m_{\text{b}})+q^2_{z}/(4m)$ into Eq. (\ref{two-body}), and  expanding Eq. (\ref{two-body}) to the order of ${\bf q}_\perp^2$, we obtain an equation satisfied by $m_{\text{b}}$ as
\begin{eqnarray}
\sum\limits_{{\bf k}}\frac{(\frac{m}{2m_{\text{b}}}-\frac{1}{4})(\mathcal{E}^2_{{\bf k},0}+\frac{4\lambda^2k^2_{\perp}}{m^2}) \mathcal{E}_{{\bf k},0}-\frac{4\lambda^4k^2_{\perp}}{m^3}\sin^2\varphi_{{\bf k}} }{(\mathcal{E}^2_{{\bf k},0}-\frac{4\lambda^2 k^2_{\perp}}{m^2})^2 \mathcal{E}_{{\bf k},0}}=0, \label{effective_mass}
\end{eqnarray}
where $\mathcal{E}_{{\bf k},0}=E_0-2\epsilon_{{\bf k}}$, and $E_0$ as a function of $1/(\lambda a_{\text{s}})$ can be obtained from Eq. (\ref{E_0}). Solving Eq. (\ref{effective_mass}) one can find a relation between $m_{\text{b}}$ and $E_0$ as \cite{Supplementray}
\begin{eqnarray}
\frac{2m}{m_{\text{b}}}=1+\frac{\lambda^2}{2m(-E_0)}\left[\frac{mE_0+\lambda^2}{-\lambda^2}
\ln\left(\frac{mE_0}{mE_0+\lambda^2}\right)-1 \right].\nonumber
\end{eqnarray}
Since $E_0<-\lambda^2/m<0$, $m_{\text{b}}$ is a monotonically decreasing as $1/(\lambda a_{\text{s}})$, as shown in the inset of Fig. \ref{Tc}. Since the bound state always exists for any $a_{\text{s}}$ \cite{Vijay}, we can discuss the BEC temperature of molecules at both sides of resonance, which is given by $T_{\text{BEC}}/T_{\text{F}}=0.218(2m/m_{\text{b}})^{2/3}$. When $1/(\lambda a_{\text{s}})\rightarrow -\infty$, $m_{\text{b}}= 4m$, which gives $T_{\text{BEC}}=0.137 T_{\text{F}}$. For a given negative $a_{\text{s}}$, $m_{\text{b}}$ decreases as $\lambda$ increases, and thus $T_{\text{BEC}}$ increases. When $1/(\lambda a_{\text{s}})\rightarrow +\infty$, $m_{\text{b}}= 2m$, which gives $T_{\text{BEC}}=0.218 T_{\text{F}}$. For a given positive $a_{\text{s}}$, $m_{\text{b}}$ increases as $\lambda$ increases, and thus $T_{\text{BEC}}$ decreases.  At resonance $a_{\text{s}}=\infty$, $m_{\text{b}}/m=2.40$ is a universal value,  and one obtains $T_{\text{BEC}}=0.193 T_{\text{F}}$. When $\lambda/k_{\text{F}}$ is large enough that the molecules become tightly bound, the actual $T_{\text{c}}$ should be very close to $T_{\text{BEC}}$, which is higher than $T_{\text{c}}=0.15T_{\text{F}}$ without SOC \cite{Tc}.

A controllable calculation of superfluid transition temperature in the entire crossover regime is a difficult task even without SOC. A widely used approximation scheme is the NSR method \cite{NSR}, in which $T_{\text{c}}$ is determined by Thouless criterion and a modified number equation
\begin{align}
\frac{m}{4\pi a_{\text{s}}}&=\frac{1}{4V}\sum\limits_{{\bf p}}\left[\frac{f_{{\bf p},+}}{(\xi_{{\bf p},+}-\mu)}+\frac{f_{{\bf p},-}}{(\xi_{{\bf p},-}-\mu)}-\frac{2}{\epsilon_{{\bf p}}}\right],\label{Thouless}\\
n&=n_{\text{fluc}}+\frac{1}{V}\sum\limits_{{\bf p}}\left[1-\frac{f_{{\bf p},+}}{2}-\frac{f_{{\bf p},-}}{2}\right].\label{numEfluc}
\end{align}
where $f_{{\bf p},\pm}=\tanh[(\xi_{{\bf p},\pm}-\mu)/(2k_{\rm B}T_{\text{c}})]$. The number equation contains the contributions from free fermions and non-condensed bosonic pairs $n_{\text{fluc}}$. In the NSR approach, $n_{\text{fluc}}$ can be obtained from diagrammatic calculations. However, such a calculation becomes much involved in the presence of SOC, and we leave it for future investigations. Here, as a rough estimation, we interpolate $T_{\text{c}}$ between $T_{\text{BCS}}$ and $T_{\text{BEC}}$ by making the approximation $n_{\text{fluc}}={1\over V}\sum_{{\bf p}}1/[e^{((p^2_{\perp}/(2m_{\text{b}})+p^2_{z}/(4m))/(k_{\text{B}}T_{\text{c}})}-1]$. In fact, such an approximation is quite reasonable in crossover regime for $\lambda/k_{\text{F}}\gtrsim 1$, since as one can see from Fig. \ref{resonance}(b), the size of pairs is already smaller than inter-particle distance. The interpolation results are shown as the solid line in Fig \ref{Tc}, from which one can see $T_{\text{c}}$ is significantly enhanced for $1/(k_{\text{F}}a_{\text{s}})\lesssim 0$; while for $1/(k_{\text{F}}a_{\text{s}})>0$, the suppression is insignificant.

As an initial effort to understand this rich system, this work points out some basic features as summarized at the beginning with simple techniques, and leave more accurate studies with more advanced techniques for future investigations. Our predications can be verified experimentally once such a system is realized. These studies are also first step toward interesting topological phases in this system with population imbalanced in two-dimension. 

{\it Acknowledgements.} We thank Hui Hu and Han Pu for sharing the manuscript before publication. This work is supported by Tsinghua University Initiative Scientific Research Program, NSFC under Grant No. 11004118 and NKBRSFC under Grant No. 2011CB921500.

{\bf Note Added:} During preparing this paper, we became aware of three preprints, in which similar problem has been addressed \cite{Shizhong_Vijay, Zhang,Hu}. For the overlap part, our results agree with each other.

\begin{widetext}

{\it Appendix:} In this supplementary material, we present some details of calculating the density of state and solving the two-body Schr\"{o}dinger equation.

\subsection{Density of State}
From the single-particle dispersion $\xi_{{\bf p},\pm}=\epsilon_p\pm \lambda p_\perp/m$, we obtain density of state for each helicity branch as
\begin{align}
  \mathcal{N}_{+}(\xi)&={1\over V}\sum_{\bf p}\,\delta(\xi-\xi_{{\bf p},+})={m\over 2\pi^2}\left[ \sqrt{2m\xi}-{\pi\lambda\over 2} +\lambda \arctan{\lambda\over \sqrt{2m\xi}}\right]\vartheta(\xi), \\
  \mathcal{N}_{-}(\xi)&={1\over V}\sum_{\bf p}\,\delta(\xi-\xi_{{\bf p},-})={m\over 2\pi^2}\left[ \sqrt{2m\xi}+{\pi\lambda\over 2} +\lambda \arctan{\lambda\over \sqrt{2m\xi}}\right]\vartheta(\xi) + {m\lambda\over 2\pi}\vartheta(-\xi)\vartheta\big(\xi+{\lambda^2\over 2m}\big),
\end{align}
where $\vartheta(\xi)$ is the step function. Comparing to the case without SOC, total DoS $\mathcal{N}(\xi)=\mathcal{N}_-(\xi)+\mathcal{N}_+(\xi)$ is always enhanced. For $-{\lambda^2\over 2m}<\xi<0$, only $\mathcal{N}_-(\xi)$ contributes to the total DoS which is a constant independent of energy.

In a ideal Fermi gas with  particle density $n$, chemical potential at zero temperature is determined by
\begin{align}
  n = {1\over V}\sum_{\bf p} \left[ \vartheta(\mu-\xi_{{\bf p},+}) + \vartheta(\mu-\xi_{{\bf p),-}} \right] = \int_0^\mu {\rm d}\xi\, \big[\mathcal{N}_+(\xi)+\mathcal{N}_-(\xi)\big].
\end{align} 
For $n<\lambda^3/(4\pi^2)$, only the helicity minus branch is occupied, and we find $\mu={4k_{\rm F}\over 3\pi\lambda}E_{\rm F}-{\lambda^2\over 2m}$. Hence in this case, DoS at Fermi surface is increased linearly as a function of SOC strength $\lambda$,
\begin{align}
  \mathcal{N}(\mu)={\pi\lambda\over 2k_{\rm F}}\mathcal{N}_0(E_{\rm F}), \quad\qquad \Big(\lambda>\big(\tfrac{4}{3\pi}\big)^{1/3}k_{\rm F}\Big)
\end{align}
where $\mathcal{N}_0(E_{\rm F})=m k_{\rm F}/ \pi^2$ is the DoS at Fermi surface without SOC. 

\subsection{Solution of Tow-body Problem}

The two-body wave-function with a center-mass momentum $\bf q$ can be written as
 \begin{align}
  |\Psi\rangle_{\bf q} = {\sum_{\bf k}}'\left[\psi_{\uparrow\downarrow}({\bf k})c_{{{\bf q}\over 2}+{\bf k}\uparrow}^\dag c_{{{\bf q}\over 2}-{\bf k}\downarrow}^\dag +\psi_{\downarrow\uparrow}({\bf k}) c_{{{\bf q}\over 2}+{\bf k}\downarrow}^\dag c_{{{\bf q}\over 2}-{\bf k}\uparrow}^\dag+ \psi_{\uparrow\uparrow}({\bf k})c_{{{\bf q}\over 2}+{\bf k}\uparrow}^\dag c_{{{\bf q}\over 2}-{\bf k}\uparrow}^\dag + \psi_{\downarrow\downarrow}({\bf k})c_{{{\bf q}\over 2}+{\bf k}\downarrow}^\dag c_{{{\bf q}\over 2}-{\bf k}\downarrow}^\dag \right]|0\rangle, \nonumber
\end{align}
where $\sum'$ denotes a summation with ${\bf k}_z>0$. Schr\"odinger equation for two-particles interacting via a contact potential 
\begin{align}
  \big(\mathcal{H}_0+\mathcal{H}_{\rm int}\big) |\Psi\rangle_{\bf q} = E_{\bf q} |\Psi\rangle_{\bf q} \label{Sch-eq}
\end{align}
can be written explicitly as
\begin{eqnarray}
   \mathcal{E}_{\bf k,q}\psi_{\uparrow\downarrow}({\bf k}) & = & {g\over V}{\sum_{\bf k'}}'\Big[\psi_{\uparrow\downarrow}({\bf k'})-\psi_{\downarrow\uparrow}({\bf k'})\Big] +  {\lambda\over m}\Big[(\tfrac{q_x}{2}-k_x)+i(\tfrac{q_y}{2}-k_y)\Big]\psi_{\uparrow\uparrow}({\bf k}) +{\lambda\over m}\Big[(\tfrac{q_x}{2}+k_x)-i(\tfrac{q_y}{2}+k_y)\Big]\psi_{\downarrow\downarrow}({\bf k}), \nonumber \\
   \mathcal{E}_{\bf k,q}\psi_{\downarrow\uparrow}({\bf k}) & = & {g\over V}{\sum_{\bf k'}}'\Big[\psi_{\downarrow\uparrow}({\bf k'})-\psi_{\uparrow\downarrow}({\bf k'})\Big] +  {\lambda\over m}\Big[(\tfrac{q_x}{2}+k_x)+i(\tfrac{q_y}{2}+k_y)\Big]\psi_{\uparrow\uparrow}({\bf k}) +{\lambda\over m}\Big[(\tfrac{q_x}{2}-k_x)-i(\tfrac{q_y}{2}-k_y)\Big]\psi_{\downarrow\downarrow}({\bf k}), \nonumber \\
  \mathcal{E}_{\bf k,q}\psi_{\uparrow\uparrow}({\bf k}) & = & {\lambda\over m}\Big[(\tfrac{q_x}{2}-k_x)-i(\tfrac{q_y}{2}-k_y)\Big]\psi_{\uparrow\downarrow}({\bf k}) + {\lambda\over m}\Big[(\tfrac{q_x}{2}+k_x)-i(\tfrac{q_y}{2}+k_y)\Big]\psi_{\downarrow\uparrow}({\bf k}), \nonumber \\
  \mathcal{E}_{\bf k,q}\psi_{\downarrow\downarrow}({\bf k}) & = &   {\lambda\over m}\Big[(\tfrac{q_x}{2}+k_x)+i(\tfrac{q_y}{2}+k_y)\Big]\psi_{\uparrow\downarrow}({\bf k}) + {\lambda\over m}\Big[(\tfrac{q_x}{2}-k_x)+i(\tfrac{q_y}{2}-k_y)\Big]\psi_{\downarrow\uparrow}({\bf k}), \nonumber
\end{eqnarray}
with $\mathcal{E}_{\bf k,q}=E_{\bf q}-\epsilon_{{{\bf q}\over 2}+{\bf k}}-\epsilon_{{{\bf q}\over 2}-{\bf k}}$.

Introducing $\psi_s({\bf k})={1\over \sqrt{2}}[\psi_{\uparrow\downarrow}({\bf k})-\psi_{\downarrow\uparrow}({\bf k})]$, $\psi_a({\bf k})={1\over \sqrt{2}}[\psi_{\uparrow\downarrow}({\bf k})+\psi_{\downarrow\uparrow}({\bf k})]$, which are corresponding to the wave-function components $|\uparrow\downarrow-\downarrow\uparrow\rangle$ and $|\uparrow\downarrow+\downarrow\uparrow\rangle$ respectively, one can find a self-consistency equation for $\psi_s$
\begin{align}
  \Big[\mathcal{E}_{\bf k,q}-{4\lambda^2 k_\perp^2\over m^2\mathcal{E}_{\bf k,q}}-{4\lambda^4(k_xq_y-k_y q_x)^2\over m^2\mathcal{E}_{\bf k,q}(m^2\mathcal{E}_{\bf k,q}^2-\lambda^2q_\perp^2)} \Big]\psi_s({\bf k}) ={g\over V}\sum_{\bf k'}\psi_s({\bf k}').
\end{align}
Hence, the energy eigenvalue $E_{\bf q}$ is determined by
\begin{align}
  {m\over 4\pi a_{\rm s}} = {1\over V}\sum_{\bf k}\bigg[{\mathcal{E}_{\bf k,q} \over \mathcal{E}_{\bf k,q}^2-{4\over m^2}\lambda^2 k_\perp^2-{4\lambda^4k_\perp^2q_\perp^2 \sin^2\varphi_{\bf qk}\over m^2(m^2\mathcal{E}_{\bf k,q}^2-\lambda^2q_\perp^2)}} +{1\over 2\epsilon_{\bf k}}\bigg], \label{binding-eq}
\end{align}
where $\varphi_{\bf qk}=\varphi_{\bf q}-\varphi_{\bf k}$ is the in-plane angle between ${\bf q}$ and ${\bf k}$. With obtained $E_{\bf q}$, the TB wave-function can also be determined through Schr\"odinger equation.

\subsection{Bound State with Zero Center-of-Mass Momentum}

For the eigen-state with center-of-mass momentum $q=0$, the threshold energy of the scattering state is $-\lambda^2/m$, and the energy of molecular bound state is determined by
\begin{align}
  {1\over 4\pi a_{\rm s}} &= {1\over V}\sum_{\bf k} \left[ {mE_0-k^2\over (mE_0-k^2)^2-4\lambda^2 k_\perp^2} +{1\over 2m\epsilon_k}\right], \nonumber \\
  & = {1\over 8\pi} \left[2\sqrt{m|E_0|} - \lambda \ln {\sqrt{m|E_0|}+\lambda \over \sqrt{m|E_0|}-\lambda} \right],
\end{align}
Introducing the molecular binding energy $E_{\rm b}=-\lambda^2/m-E_0$, above equation can be re-written as
\begin{align}
  {1\over \lambda a_{\rm s}} = {\sqrt{mE_{\rm b}+\lambda^2}\over \lambda}- {1\over 2}\ln {\sqrt{mE_{\rm b}+\lambda^2}+\lambda\over\sqrt{mE_{\rm b}+\lambda^2}-\lambda}.
\end{align}
One can see that for any value of $a_{\rm s}$ there is always a bound state with $E_{\rm b}>0$. In the limit of $\lambda a_{\rm s}\rightarrow 0^+$, $|E_0|\gg \lambda^2/m$, we obtain
\begin{align}
  E_{\rm b}={1\over m a_{\rm s}^2},
\end{align}
which is same as the case without SOC. 
In the limit of $\lambda a_{\rm s}\rightarrow 0^-$, $|E_0|\simeq \lambda^2/m$, we obtain
\begin{align}
  E_{\rm b} = {4\lambda^2\over e^2} e^{-{2\over \lambda |a_{\rm s}|}},
\end{align}
which shows a exponential dependence on $(\lambda a_{\rm s})^{-1}$. At unitarity, $a_{\rm s}\rightarrow\infty$, we find
\begin{align}
  E_{\rm b}=0.439 {\lambda^2\over m}.
\end{align}
 
The bound state wave-function is given by
\begin{align}
\psi_{\uparrow\downarrow}({\bf k})&=-\psi_{\downarrow\uparrow}({\bf k})={1\over \sqrt{\mathcal{C}'}} \left[{1\over E_0-\xi_{{\bf k},+}} + {1\over E_0-\xi_{{\bf k},-}} \right], \\
  \psi_{\uparrow\uparrow}({\bf k})&=-\psi_{\downarrow\downarrow}^*({\bf k})={-1\over \sqrt{\mathcal{C}'}} \left[{1\over E_0-\xi_{{\bf k},+}} - {1\over E_0-\xi_{{\bf k},-}} \right]e^{-i\varphi_{\bf k}},
\end{align}
where $\mathcal{C}'$ is the normalization coefficient. The fact $\psi_{a}({\bf k})=0$ implies that for the bound state with $q=0$ the triplet component of $|\uparrow\downarrow+\downarrow\uparrow\rangle$ vanishes.

\subsection{Effective Mass of Molecule} 

For the bound state with a finite center-of-mass momentum, the eigen-energy can be written as $E_{\bf q}=E_0+q_\perp^2/(2m_{\rm b})+q_z^2/(4m)$ if $q$ is small enough. Substituting this dispersion in Eq. (\ref{binding-eq}) and expanding to the order of $q_\perp^2$, we find
\begin{align}
\big({2m\over m_{\rm b}}-1\big) \sum_{\bf k} {\mathcal{E}_{{\bf k},0}^2+{4\over m^2}\lambda^2 k_\perp^2 \over \big(\mathcal{E}_{{\bf k},0}^2-{4\over m^2}\lambda^2k_\perp^2\big)^2} = \sum_{\bf k} {16\lambda^4 k_\perp^2 \sin^2\varphi_{\bf k} \over  m^3\mathcal{E}_{{\bf k},0}\big(\mathcal{E}_{{\bf k},0}^2-{4\over m^2}\lambda^2k_\perp^2\big)^2} \label{mass_eq},
\end{align}
where $\mathcal{E}_{{\bf k},0}=E_0-k^2/m$. The integral on the l.h.s. above can be computed straightforwardly as
\begin{align}
  {1\over V} \sum_{\bf k} {\mathcal{E}_{{\bf k},0}^2+{4\over m^2}\lambda^2 k_\perp^2 \over \big(\mathcal{E}_{{\bf k},0}^2-{4\over m^2}\lambda^2k_\perp^2\big)^2}
  & = {1\over 2V} \sum_{\bf k} \left[ {1\over (\mathcal{E}_{{\bf k},0}-{2\over m}\lambda k_\perp)^2} + {1\over (\mathcal{E}_{{\bf k},0}+{2\over m}\lambda k_\perp)^2} \right] \nonumber \\
  & = {m^2\over 8\pi^2} \int_0^\infty {\rm d}k_\perp \int_{-\infty}^\infty {\rm d}k_z\, \left[{k_\perp \over (mE_0-k_\perp^2 - k_z^2 - 2\lambda k_\perp)^2} + {k_\perp \over (mE_0-k_\perp^2 - k_z^2 + 2\lambda k_\perp)^2} \right] \nonumber \\
  & = {m^2\over 16\pi} \int_0^\infty {\rm d}k_\perp  \left[{k_\perp \over (k_\perp^2 +2\lambda k_\perp-mE_0)^{3/2}} + {k_\perp \over (k_\perp^2 -2\lambda k_\perp-mE_0)^{3/2}} \right] \nonumber \\
  & = {m^2 \over 8\pi} {\sqrt{m|E_0|} \over m|E_0|-\lambda^2},
\end{align}
and the integral on the r.h.s of Eq. (\ref{mass_eq}) can be computed as
\begin{align}
  &{1\over V}\sum_{\bf k} {16 \lambda^4 k_\perp^2 \sin^2\varphi_{\bf k} \over m^3 \mathcal{E}_{{\bf k},0}\big(\mathcal{E}_{{\bf k},0}^2-{4\over m^2}\lambda^2k_\perp^2\big)^2} \nonumber \\
  = & {1\over V}\sum_{\bf k} {\lambda \sin^2\varphi_{\bf k} \over 2 k_\perp} \left[{1\over (\mathcal{E}_{{\bf k},0}-{2\over m}\lambda k_\perp)^2} - {1\over (\mathcal{E}_{{\bf k},0}+{2\over m}\lambda k_\perp)^2} - {m\over \lambda k_\perp}\Big( {1\over \mathcal{E}_{{\bf k},0}-{2\over m}\lambda k_\perp}+ {1\over \mathcal{E}_{{\bf k},0}+{2\over m}\lambda k_\perp}-{2\over\mathcal{E}_{{\bf k},0}}\Big) \right] \nonumber \\
  = & {m^2\lambda\over 16\pi^2} \int_0^\infty {\rm d}k_\perp \int_{-\infty}^\infty {\rm d}k_z\, \bigg[ {1\over (mE_0-k_\perp^2-k_z^2-2\lambda k_\perp)^2} - {1\over (mE_0-k_\perp^2-k_z^2+2\lambda k_\perp)^2} \nonumber \\
  & \qquad \qquad \qquad \qquad \qquad \quad -{1\over \lambda k_\perp}\Big({1\over mE_0-k_\perp^2-k_z^2-2\lambda k_\perp}+ {1\over mE_0-k_\perp^2-k_z^2+2\lambda k_\perp} - {2\over mE_0-k_\perp^2-k_z^2}\Big) \bigg] \nonumber \\
  = & {m^2\lambda\over 32\pi} \int_0^\infty {\rm d}k_\perp \bigg[ {1\over (k_\perp^2+2\lambda k_\perp - mE_0)^{3/2}} -  {1\over (k_\perp^2-2\lambda k_\perp - mE_0)^{3/2}} \nonumber \\
  & \qquad \qquad \qquad \qquad \qquad \quad -{2\over \lambda k_\perp} \Big( {1\over \sqrt{k_\perp^2+2\lambda k_\perp-mE_0}} + {1\over \sqrt{k_\perp^2-2\lambda k_\perp-mE_0}} -{2\over \sqrt{k_\perp^2-mE_0}} \Big) \bigg] \nonumber \\
  = & {m^2\over 16\pi\sqrt{m|E_0|}} \left[\ln{m|E_0|\over m|E_0|-\lambda^2} - {\lambda^2 \over m|E_0|-\lambda^2} \right].
\end{align}
Thus the in-plane molecular effective mass $m_{\rm b}$ is given by
\begin{align}
  {2m\over m_{\rm b}} &= 1 - {1\over 2} \left[{m|E_0|-\lambda^2 \over m|E_0|}\ln{m|E_0|-\lambda^2\over m|E_0|}+{\lambda^2\over m|E_0|}\right] \nonumber \\
  &= 1-{1\over 2} \left[{ mE_{\rm b}\over mE_{\rm b}+\lambda^2}\ln{ mE_{\rm b}\over mE_{\rm b}+\lambda^2}+{\lambda^2\over mE_{\rm b}+\lambda^2}\right].
\end{align}
One can see that $m_{\rm b}$ is always larger than $2m$ for non-zero $\lambda$.
In the limit of $\lambda a_{\rm s}\rightarrow 0^+$, $E_{\rm b}\gg \lambda^2/m$, we find $$m_{\rm b}=2m.$$
In the limit of $\lambda a_{\rm s}\rightarrow 0^-$, $E_{\rm b}\ll \lambda^2/m$, we find $$m_{\rm b}=4m.$$
At unitarity, $E_{\rm b}=0.439\lambda^2/m$, we obtain $$m_{\rm b}=2.40m.$$

\end{widetext}

\end{document}